\pdfoutput=1

\documentclass[11pt]{article}

\usepackage[final]{acl}

\usepackage{times}
\usepackage{latexsym}

\usepackage[T1]{fontenc}

\usepackage[utf8]{inputenc}

\usepackage{microtype}

\usepackage{inconsolata}

\usepackage{graphicx}

\usepackage{xcolor}
\usepackage{multirow}
\usepackage{multicol}
\usepackage{array}
\usepackage{booktabs,threeparttable}
\usepackage{makecell}
\usepackage{amssymb}
\usepackage{placeins}

\title{Batched Self-Consistency Improves LLM Relevance Assessment and Ranking} 

\author{
  \textbf{Anton Korikov\textsuperscript{1,2}},
  \textbf{Pan Du\textsuperscript{2}},
  \textbf{Scott Sanner\textsuperscript{1}},
  \textbf{Navid Rekabsaz\textsuperscript{2}}
\\
  \textsuperscript{1}University of Toronto\\
  \textsuperscript{2}Thomson Reuters Labs
\\
  \small{
    \textbf{Correspondence:} \href{mailto:}{anton.korikov@mail.utoronto.ca}
  }
\\
\texttt{\{korikov,ssanner\}@mie.utoronto.ca} 
\\
\texttt{\{pan.du,navid.rekabsaz\}@thomsonreuters.com} 
}

\usepackage{fancyhdr}
\fancypagestyle{firstpage}{
  \fancyhf{} 
  \fancyfoot[C]{\small \textit{EMNLP '25, Suzhou, China, Nov 4 -- 9, 2025}}
}
\begin{document}
\maketitle
\begin{abstract}
LLM query-passage relevance assessment is typically studied using a one-by-one pointwise (PW) strategy where each LLM call judges one passage at a time. However, this strategy requires as many LLM calls as there are passages while also preventing information sharing between passages. We thus hypothesize that batched PW methods, which evaluate multiple passages per LLM call, can improve not only efficiency but also judgment quality -- by enabling content from multiple passages to be seen jointly. Moreover, batched PW methods may be better suited to harness the test-time scaling benefits of self-consistency -— the ensembling technique of repeating (potentially perturbed) LLM tasks in parallel and aggregating results -— since batching can naturally enable prompt diversification through varied batch permutations and compositions to create more robust ensembles. We evaluate several batched PW methods against one-by-one PW and listwise ranking baselines on LLM relevance assessment and ranking tasks, using three passage retrieval datasets and GPT-4o, Claude Sonnet 3, and Amazon Nova Pro. We show that batching can greatly amplify self-consistency benefits, making batched PW methods achieve the best performance while often reducing latency by an order of magnitude or more compared to one-by-one PW methods. For instance, on legal search, batched PW ranking with GPT-4o improves from 43.8\% to 51.3\% NDCG@10 when using 1 vs. 15 self-consistency calls, compared to one-by-one PW ranking improving from 44.9\% to 46.8\% and being 15.3x slower.
\end{abstract}


\section{Introduction}
\begin{figure*}
    \centering
    \includegraphics[width=\linewidth]{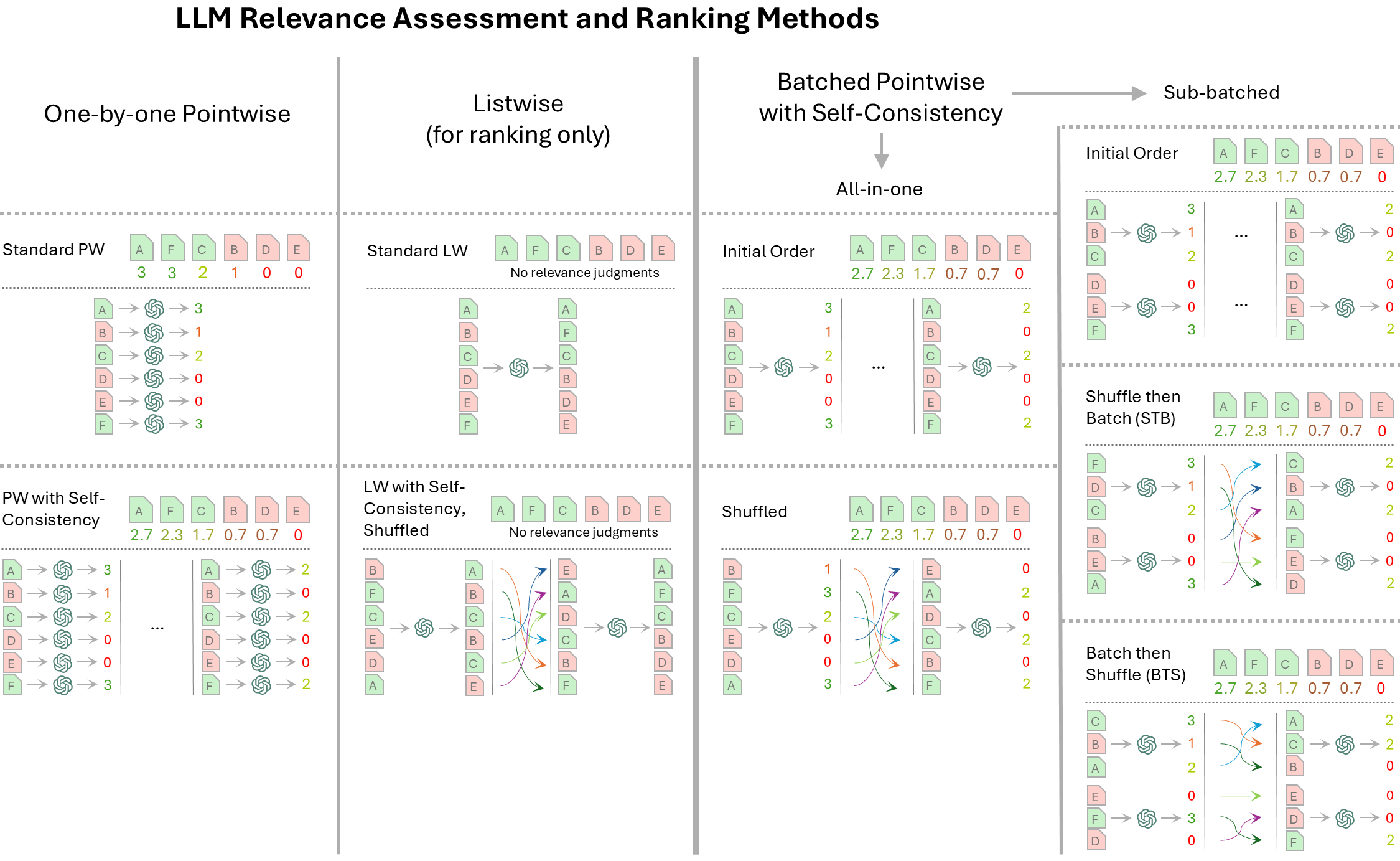}
    \caption{
    One-by-one PW methods (first pane) evaluate each candidate in a separate LLM call. LW methods (second pane) prompt the LLM to rerank the candidate list but do not produce relevance judgments. 
    All-in-one batched PW methods (third pane) evaluate \textit{all} candidates in each call while sub-batched PW methods (right pane) select a subset. Self-consistency calls can repeat identical LLM calls, or use various candidate permutations (LW and batched PW only), and/or different candidate subsets (sub-batched PW only).}
    \label{fig:main}
\end{figure*}

Given a query and a set of candidate passages, LLM query-passage relevance assessment has become a foundational task for agentic artificial intelligence (AI) and retrieval-augmented generation (RAG) \cite{wang2024survey}. Existing work has focused on a one-by-one pointwise (PW) scoring strategy where each LLM call judges one passage at a time against the query \cite{sachan2022improving, zhuang2023beyond, thomas2024large,upadhyay2024umbrela,tornberg2024best, ma2024fine}. However, this strategy does not allow content from multiple candidate passages to be seen jointly, and can be computationally expensive as it requires as many LLM calls as there are passages.
In this work, we hypothesize that batched PW scoring -- evaluating multiple candidates in a single LLM call -- can improve both judgment quality and computational efficiency, since content from multiple passages can be seen jointly while requiring fewer LLM calls.

Further, we ask how LLM relevance assessment methods can leverage 
the test-time scaling benefits of 
self-consistency  \cite{wangself} -- a parallelizable ensembling technique where the same LLM task is repeated, potentially with perturbations, and the results are aggregated. To collect multiple scores per passage, one-by-one methods are generally limited to repeating identical LLM calls, relying only on LLM stochasticity to produce different scores. Batched PW methods, however, can naturally perturb the contexts in which scores for a given candidate are generated by using different co-candidate subsets and permutations across batches, as shown in Figure \ref{fig:main}. Moreover, autoregressively generating a sequence of scores, as opposed to only one score at a time, could diversify scoring further. We thus conjecture that batching may create more robust self-consistency ensembles that exhibit stronger test time scaling due to more diversification across ensemble components. 

We conduct experiments on the tasks of passage relevance assessment (i.e., predicting the relevance label of a passage) and ranking, using three passage retrieval datasets and three LLMs (GPT-4o, Clause Sonnet 3, and Amazon Nova Pro). We compare batched PW methods against one-by-one PW and listwise (LW) \cite{ma2023zero, tang2024found, hou2024large} baselines -- for each method, we also investigate the effects of self-consistency for ensemble sizes from 1-15 components. Our contributions include: 

\begin{itemize}
    \item We show that batched PW methods can improve not only efficiency but also judgment quality by enabling content from multiple passages to be seen jointly.
    \item We propose several batched PW strategies to diversify self-consistency ensembles via various candidate subsets and permutations.
    \item We find that batching produces stronger self-consistency ensembles with faster test-time scaling. For instance, on legal search, batched PW ranking with GPT-4o improves from 43.8\% to 51.3\% NDCG@10 when using 1 vs. 15 self-consistency calls, compared to one-by-one PW ranking which improves from 44.9\% to 46.8\%.
    \item We show that batching can reduce latency by an order of magnitude or more when the number of available LLMs does not exceed the number of ensemble components.
\end{itemize}

\section{Related Work}
We briefly review existing LLM and non-LLM text relevance assessment and ranking techniques as well as emerging work on LLM self-consistency. 

\subsection{Text Relevance Assessment and Ranking}
\label{sec:rel_work_ranking}
Given a query $q$, both non-generative and generative approaches are widely used to rank or score candidate text spans in some collection $\{p_1,\cdots,p_D\}$ based on relevance to $q$. Non-neural, sparse methods such as TF-IDF \cite{salton1975vector} and its probabilistic variant BM25 \cite{robertson2009probabilistic} rely on syntactic token matches, limiting their ability to capture semantic similarity. Encoder-only LLM methods broadly include bi-encoders \cite{izacard2021unsupervised, gao2021condenser} which score using a similarity function (e.g. cosine similarity) between separately embedded queries and passages, and cross-encoders \cite{nogueira2019passage, zhuang2023rankt5} which jointly embed queries and passages to predict a relevance score.
While these foundational methods are critical for retrieving initial candidate lists from large corpora, none of them are able to benefit from self-consistency since they do not use prompting or stochastic decoding.

Generative LLM-based methods typically consist of PW, pairwise, and LW strategies. Standard one-by-one PW and LW techniques are baselines discussed further in sections \ref{sec:obo_pw} and \ref{sec:lw}, respectively.  Pairwise rankers \cite{qin2024large,liu2024lost} ask an LLM which passage out of a pair $(p_i,p_j)$ is more relevant, but need a quadratic number of LLM calls relative to the number of candidates. There are also bubble-sort sliding-window LW LLM rankers \cite{ma2023zero,sun2023chatgpt, pradeep2023rankvicuna, pradeep2023rankzephyr} which achieve beneficial test-time scaling effects by using  multiple sequential LLM calls to iteratively rank overlapping intervals of the initial list. We leave an exploration of sequential test-time scaling for batched PW methods for future work, focusing instead on fully parallelizable self-consistency ensembles.








\subsection{LLM Self-Consistency}



LLM self-consistency \cite{wangself} is a parallelizable ensembling technique that involves performing the same LLM task multiple times and  aggregating the outputs. The variation between outputs can occur not only because of stochastic response generation but also from perturbations to the prompt. Many approaches exist \cite{li2024more}, including diversifying few-shot examples across self-consistency calls \cite{lu2022fantastically}, using token entropy to weigh or prune components \cite{kang2025scalable,fu2025deep}, creating ensembles of heterogenous LLMs \cite{wan2024knowledge}, and recent work on LW ranking that uses different passage perturbations across calls \cite{tang2024found, hou2024large}, as discussed further in Section \ref{sec:lw}.    


\section{Methodology}
Figure \ref{fig:main} outlines the methods we study for LLM relevance assessment and ranking, all of which take as input a query $q$ and an initial list of $D$ candidate passages $L^q = [p_1,...,p_D]$. All methods rerank the initial list, but only PW methods generate relevance predictions.


\paragraph{PW Relevance Labels:} In all PW methods, each LLM call generates a relevance score $s_{q,p} \in \mathbb{R}$ between $q$ and each passage $p \in L^q$. For ranking, passages are sorted in descending score order with ties broken using the order of the initial list $L^q$. We use the following 0-3 scale from the UMBRELLA open-source Bing prompt \cite{upadhyay2024umbrela}:
\begin{itemize}
    \item \textbf{3:} The passage is dedicated to the query and contains the exact answer.
    \item \textbf{2:} The passage has some answer for the query, but the answer may be a bit unclear, or hidden amongst extraneous information.
    \item \textbf{1:} The passage seems related to the query but does not answer it.
    \item \textbf{0:} The passage has nothing to do with the query.
\end{itemize}

Our full prompt is shown in Appendix \ref{sec:appendix_prompts}.

\paragraph{Self-Consistency} All our self-consistency methods include each passage in exactly $m$ LLM calls. In our PW self-consistency methods, each passage thus receives $m$ scores $\{s^1_{q,p},\cdots,s^m_{q,p}\}$, which are aggregated into a final score $s_{q,p}$ by taking the mean.\footnote{For ranking with integer 0-3 scores, mean aggregation reduces the number of  ties compared to majority voting.} Similarly, our LW self-consistency methods aggregate $m$ output lists of length $D$ by minimizing Kendall-Tau distance, as described further in Sec. \ref{sec:lw}.   


\subsection{Baselines}
\label{sec:baseline}

\subsubsection{One-by-one Pointwise Methods}
\label{sec:obo_pw}
The first pane of Figure \ref{fig:main} shows one-by-one PW methods \cite{sachan2022improving,zhuang2023beyond,thomas2024large,upadhyay2024umbrela,tornberg2024best, ma2024fine}, where each passage is scored against the query in a separate LLM call. While evaluating one candidate at a time avoids inducing candidate position biases, it also prevents the LLM from seeing potentially helpful context in other passages. Further, each passage score requires a separate LLM call, which can lead to a very large number of calls. Finally, when self-consistency ensembling is used, one-by-one methods are typically limited to repeating identical LLM calls, relying only on LLM stochasticity to generate different responses. In contrast, methods which include more than one candidate can naturally diversify prompts across an ensemble by varying co-candidate selections and permutations.



\subsubsection{Listwise Ranking Methods}
\label{sec:lw}
LW ranking methods \cite{ma2023zero, tang2024found, hou2024large, reddy2024first} are shown in the second pane of Figure \ref{fig:main}.

\paragraph{Standard LW:} Standard LW ranking instructs an LLM to rerank an input passage list in order of relevance with respect to $q$, with our LW prompt shown in Appendix \ref{sec:appendix_prompts}. While the LLM sees all available passage context during a single inference, no absolute relevance judgments are produced.

\paragraph{LW with Self-Consistency:} LW ranking with self-consistency \cite{tang2024found, hou2024large} involves $m$ reranking calls followed by a rank aggregation of $m$ output lists, which can be done by minimizing the Kendall-Tau distance with the $m$ lists. Multiple exact and approximate aggregation techniques exist, with our experiments using the exact Kemeny rank aggregation linear program (LP) of \citeauthor{tang2024found}\footnote{https://github.com/castorini/perm-sc} We test two LW self-consistency variants: 
\begin{enumerate}
    \item \textbf{Initial Order:} Each of $m$ LLM calls is identical and maintains the initial input list order.
    \item \textbf{Shuffled:} The  passage list is fully shuffled before each LLM call.
\end{enumerate}

\subsection{Batched PW Methods}
The right half of Figure \ref{fig:main} shows batched PW methods in which multiple passages are jointly scored in each LLM call. This allows an LLM to see context from multiple passages at the same time and reduces the total number of LLM calls required compared to one-by-one methods. We also study the effects of diversifying passage subsets and permutations across self-consistency ensemble prompts through several batching strategies, including all-in-one batching 
(c.f. Sec. \ref{sec:aio}) and sub-batching
(c.f. Sec. \ref{sec:sub_b}). Note that autoregressively generating a sequence of scores, as opposed to only one score at a time as in one-by-one methods, could diversify scores between ensemble components further.

\subsubsection{All-in-one}
\label{sec:aio}
All-in-one PW methods prompt the LLM to score all passages in a single batch,\footnote{The LLM context window must be large enough to fit all  passages, otherwise sub-batching (c.f. Sec \ref{sec:sub_b}) is required.} maximizing the context available to the model but also presenting it with the most complex task. We test two passage ordering strategies:
\begin{enumerate}
    \item \textbf{Initial Order:} The initial list order is kept, giving $m$ identical self-consistency calls.
    \item \textbf{Shuffled}: The passage list is fully shuffled before each self-consistency LLM call, giving $m$ calls with $m$ random passage permutations.
\end{enumerate}

\begin{figure*}[t]
    \centering
    \includegraphics[width=\linewidth]{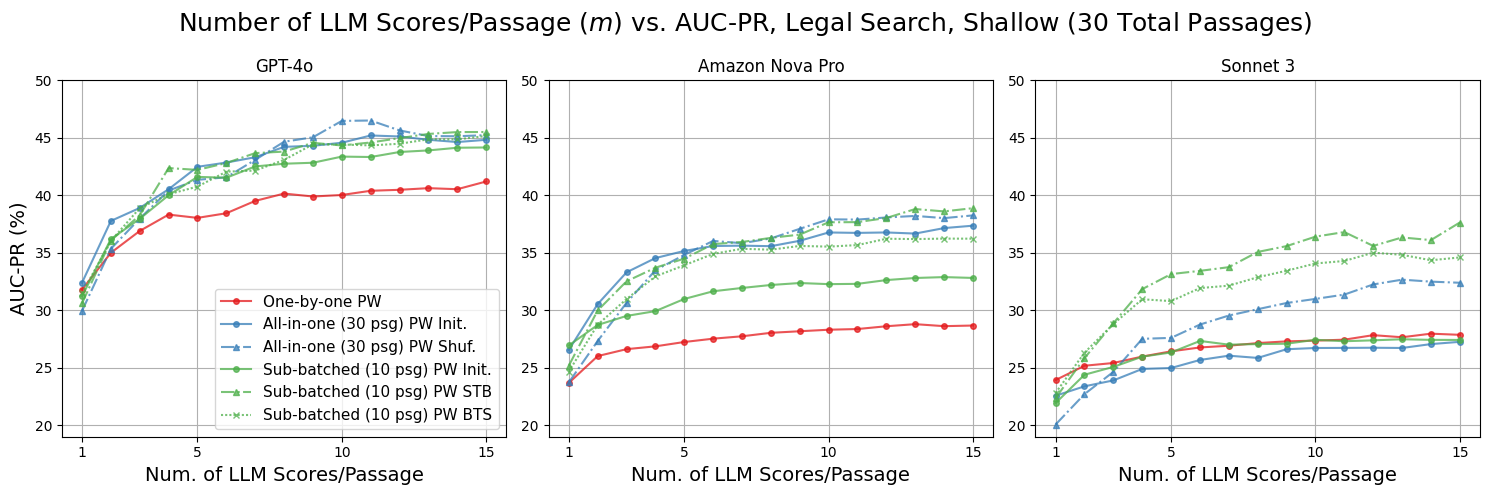}
    \caption{Effect of increasing self-consistency calls/passage ($m$) on PW relevance assessment quality (Legal Search, Shallow). For all LLMs at $m = 1$ (no self-consistency), one-by-one PW is competitive, but by $m = 15$ it underperforms batched PW by  5-10\% AUC-PR. This shows that batching amplifies the benefits of self-consistency, likely because it 
   allows passages to be seen jointly and creates more diverse self-consistency ensembles.}
    \label{fig:n_scores_shall}
\end{figure*}

\begin{figure*}[t]
    \centering
    \includegraphics[width=\linewidth]{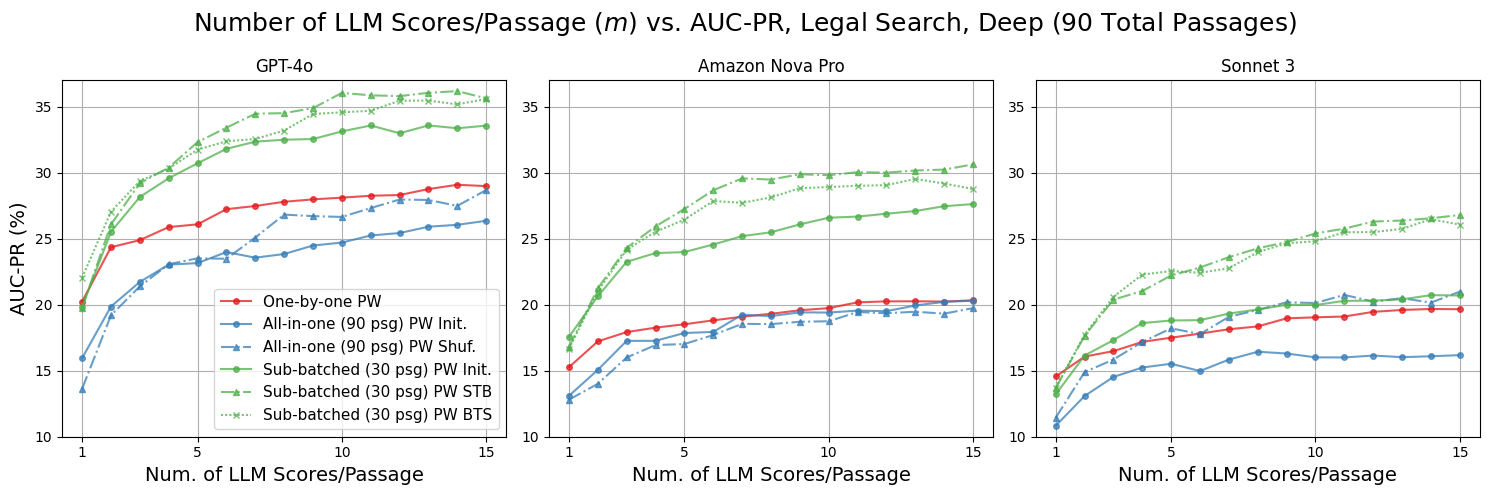}
    \caption{Effect of increasing $m$ on AUC-PR for Deep Legal Search (90 Total Psgs), showing that overly large batch sizes can harmful. The sub-batched methods (30 psgs/batch) perform very well, with the shuffling variants (STB and BTS) doing best. The large batch (90 psgs/batch) all-in-one methods perform poorly for this range of $m$, addressed further in RQ2.}
    \label{fig:n_scores_deep}
\end{figure*}

\subsubsection{Sub-batched}
\label{sec:sub_b}
Sub-batching methods select subsets of passages for each batch, providing less context than all-in-one batching but also asking the LLM to generate fewer scores. We consider the following context selection and permutation strategies:

\begin{figure*}
    \centering
    \includegraphics[width=\linewidth]{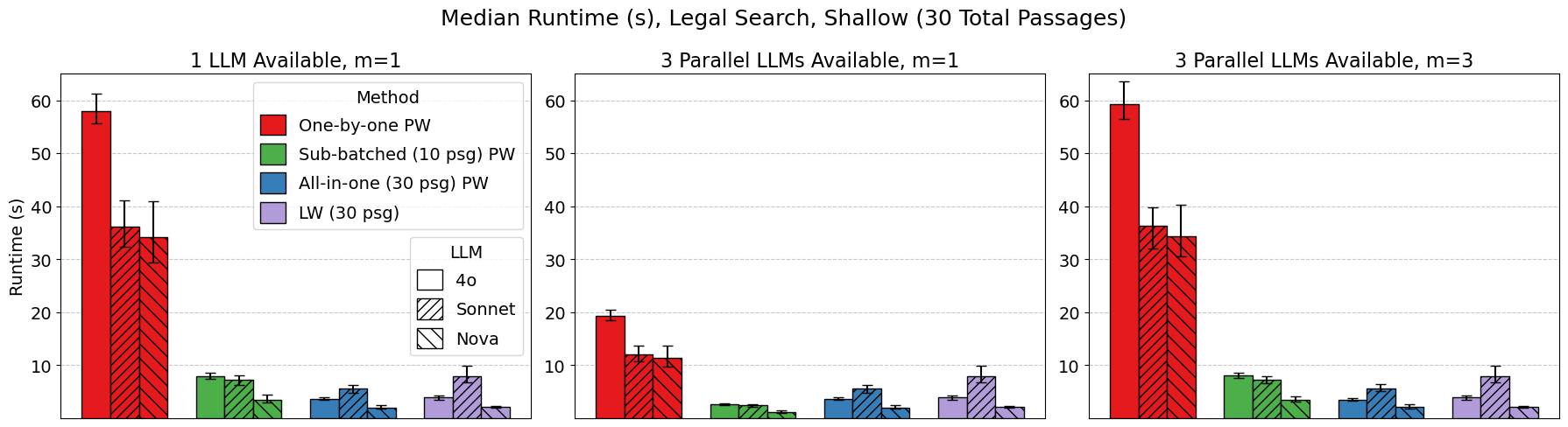}
    \caption{Median per-query runtimes for Legal Search, Shallow (30 total passages), with error bars showing the IQR. The 1st and 3rd plots show 6-17x speedups from batching. If the number of parallel LLMs available is $> m$, as in the 2nd plot, smaller batches may allow more parallelization. However, if only a few  LLMs are available, processing multiple passages per call is typically much faster, especially with self-consistency ensembling.}
    \label{fig:runtimes}
\end{figure*}

\begin{table*}[t]
\centering
\renewcommand{\arraystretch}{1.25}
\setlength{\tabcolsep}{3pt} 
\scriptsize
\begin{tabular}{|c|c|c|c||c|c|c||c|c|c||c|c|c|}
\hline
\multicolumn{1}{|c|}{} & & & &
\multicolumn{3}{c||}{\textbf{Legal Search}} &
\multicolumn{3}{c||}{\textbf{DL-19}} &
\multicolumn{3}{c|}{\textbf{Covid}} \\
\cline{5-13}
\multicolumn{1}{|c|}{} &
\begin{tabular}[c]{@{}c@{}}\textbf{PW/}\\\textbf{LW}\end{tabular} &
\begin{tabular}[c]{@{}c@{}}\textbf{Psgs/}\\\textbf{Batch}\end{tabular} &
\begin{tabular}[c]{@{}c@{}}\textbf{Batches/}\\\textbf{Query}\end{tabular} &
\textbf{4o} & \textbf{Sonnet} & \textbf{Nova} &
\textbf{4o} & \textbf{Sonnet} & \textbf{Nova} &
\textbf{4o} & \textbf{Sonnet} & \textbf{Nova} \\
\Xhline{4\arrayrulewidth}

\multirow{4}{*}{\rotatebox{90}{Shallow (30)}}
& PW & 1  & 30 & 469    & 521    & 470    & 272   & 312   & 266   & 522    & 596    & 522    \\ \cline{2-13}
& PW & 10 & 3  & 2,956  & 3,212  & 3,032  & 1,074 & 1,208 & 1,082 & 3,502  & 3,968  & 3,569  \\ \cline{2-13}
& PW & 30 & 1  & 8,455  & 9,147  & 8,696  & 2,853 & 3,197 & 2,903 & 10,125 & 11,468 & 10,351 \\ \cline{2-13}
& LW & 30 & 1  & 8,443  & 9,133  & 8,685  & 2,841 & 3,183 & 2,892 & 10,113 & 11,454 & 10,340 \\
\Xhline{4\arrayrulewidth}

\multirow{4}{*}{\rotatebox{90}{Deep (90)}}
& PW & 1  & 90 & 449    & 500    & 449    & 271   & 310   & 265   & 525    & 599    & 526    \\ \cline{2-13}
& PW & 30 & 3  & 7,877  & 8,527  & 8,106  & 2,825 & 3,161 & 2,873 & 10,199 & 11,554 & 10,443 \\ \cline{2-13}
& PW & 90 & 1  & 23,179 & 25,063 & 23,892 & 8,094 & 9,045 & 8,267 & 30,268 & 34,286 & 31,136 \\ \cline{2-13}
& LW & 90 & 1  & 23,167 & 25,049 & 23,881 & 8,082 & 9,031 & 8,256 & 30,256 & 34,272 & 31,125 \\ \hline
\end{tabular}
\caption{Mean input token counts per LLM call.}
\label{tab:tokens}
\end{table*}

\begin{enumerate}
    \item \textbf{Initial Order:} To create $B$ batches, the initial list is partitioned into $B$ non-overlapping intervals while maintaining its order, e.g., $[p_1,\cdots, p_{30}]$ $\rightarrow$ $[p_1\cdots,p_{10}],$ $[p_{11}\cdots,p_{20}],$ $[p_{21},\cdots,p_{30}]$, where $p_i$ is the $i$'th passage in the initial list. Self-consistency calls for each sub-batch are identical and repeat the the same biases,
    giving this sub-batching strategy the least ensemble diversity.
    \item \textbf{Shuffled-then-Batched (STB):} The passage list is fully shuffled and then split into $B$ batches before each LLM call, creating the most diversity in passage subsets and permutations across self-consistency ensembles.
    \item \textbf{Batched-then-Shuffled (BTS):} The initial list is first divided into $B$ intervals, as in Initial Order Sub-batching above, and then each batch $b$ is fully shuffled before each LLM call. While BTS attempts to mitigate position bias through shuffling, the passage mixture for a given batch remains constant across self-consistency calls, making its LLM sampling less diverse than STB.   
\end{enumerate}

\section{Experimental Setup}
We evaluate the effects of batching and self-consistency on the relevance assessment and ranking abilities of three LLMs, namely GPT-4o (128K), Sonnet 3 (200K), and Amazon Nova Pro (300K) at temperature 1 across three passage retrieval datasets, releasing code.\footnote{https://anonymous.4open.science/r/batched-sc-emnlp/} 

\paragraph{Self-consistency:} We vary the number of self-consistency calls per passage, $m$, from 1 (i.e., no self-consistency) to 15. For each query, each passage appears in exactly $m$ calls.

\subsection{Passage Relevance Assessment and Ranking Tasks}
Each passage retrieval task contains a set of queries $\mathcal{Q}$, a corpus of passages $\mathcal{D}$, and a relevance label $y_{q,p} \in \mathbb{R}$ for each query-passage pair.
For each $q$, we also have an list of $D$ passages $L^q = [p_1,\cdots,p_D]$ returned by some initial retrieval algorithm (e.g., BM25, dense retrieval).

\paragraph{Metrics:} We use NDCG@10 to evaluate LLM ranking quality. We treat relevance assessment as a binary classification task, converting each LLM score $s_{q,p} \in [0,3]$ to a relevance probability $p(\hat{y}_{q,p} = 1) = s_{q,p}/3$, and using the area under the precision recall curve (AUC-PR) for evaluation.\footnote{AUC-PR is preferred to AUC-ROC for imbalanced data \cite{saito2015precision}, and IR datasets are highly imbalanced.}

\paragraph{Shallow vs. Deep Search} To study the effect of the total number of passages evaluated, we test a short ($D = 30$) and long ($D = 90$) initial list length in what we call \textit{shallow} and \textit{deep} search, respectively. All sub-batching methods split the initial list into three equal sized batches (i.e., 10 and 30 passages per batch for shallow and deep search, respectively).

\subsection{Datasets}
Our evaluation includes two well-known open-source passage retrieval datasets, TREC DL-19 \cite{craswell2019overview} 
and TREC Covid \cite{voorhees2021trec} with BM25 used to retrieve the initial passage list $L^q$ for each $q$. We also test a third, closed-source dataset from Thomson Reuters which we call Legal Search. Legal Search comprises 100 legal queries and paragraph-size passages chunked from 100,000 proprietary legal documents and uses the output of a multi-stage industrial retrieval pipeline to retrieve $L^q$. 

\paragraph{Token Usage}
Table \ref{tab:tokens} shows the mean input token counts per LLM call for each method. The longest resulting input context (90 TREC Covid passages) is 30K-34K tokens, which is well within context limits for all three LLMs.


\section{Experimental Results}



\renewcommand{\arraystretch}{1.3} 
\begin{table*}[t]
\scriptsize
\centering
\begin{tabular}{|c|c|c||c|c|c||c|c|c||c|c|c||}
\cline{2-12}
\multicolumn{1}{c|}{} & & & \multicolumn{3}{c||}{\textbf{Legal Search}} & \multicolumn{3}{c||}{\textbf{DL-19}} & \multicolumn{3}{c||}{\textbf{Covid}} \\
\cline{4-12}
\multicolumn{1}{c|}{} & &  & \textbf{GPT-4o} & \textbf{Sonnet} & \textbf{Nova} & \textbf{GPT-4o} & \textbf{Sonnet} & \textbf{Nova} & \textbf{GPT-4o} & \textbf{Sonnet} & \textbf{Nova} \\
\multicolumn{1}{c|}{} & \textbf{Psg/} & \textbf{Psg} & \textbf{AUC\textsuperscript{1}/} & \textbf{AUC\textsuperscript{1}/} & \textbf{AUC\textsuperscript{1}/} & \textbf{AUC\textsuperscript{1}/} & \textbf{AUC\textsuperscript{1}/} & \textbf{AUC\textsuperscript{1}/} & \textbf{AUC\textsuperscript{1}/} & \textbf{AUC\textsuperscript{1}/} & \textbf{AUC\textsuperscript{1}/} \\
\multicolumn{1}{c|}{} & \textbf{Batch} & \textbf{Order} & \textbf{AUC\textsuperscript{15}} & \textbf{AUC\textsuperscript{15}} & \textbf{AUC\textsuperscript{15}} & \textbf{AUC\textsuperscript{15}} & \textbf{AUC\textsuperscript{15}} & \textbf{AUC\textsuperscript{15}} & \textbf{AUC\textsuperscript{15}} & \textbf{AUC\textsuperscript{15}} & \textbf{AUC\textsuperscript{15}} \\
\Xhline{4\arrayrulewidth}
\multirow{6}{*}{\rotatebox{90}{Shallow (30 psg)}}
& 1 & --  & \textbf{32} / 41 & \textbf{24} / 28 & 24 / 29 & 39 / 52 & \textbf{27} / 31 & 29 / 33 & \textbf{70} / 78 & \textbf{68} / 73 & \textbf{68} / 77 \\
\cline{3-12}
\Xcline{2-12}{2.5\arrayrulewidth}
& 10 & Init.  & 31 / 44 & 22 / 27 & \textbf{27} / 33 & 34 / 51 & 25 / 33 & \textbf{32} / 40 & 70 / 77 & 63 / 70 & 68 / 74 \\
\cline{3-12}
 & (sub-batch) & STB  & 31 / \textbf{46} & 22 / \textbf{38} & 25 / \textbf{39} & 39 / 57 & 25 / \textbf{41} & 30 / 47 & 69 / \textbf{80} & 65 / \textbf{77} & 68 / \textbf{79} \\
\cline{3-12}
& & BTS & 31 / 45 & 23 / 35 & 25 / 36 & 39 / 55 & 25 / \textbf{41} & \textbf{32} / 46 & 69 / 79 & 63 / 75 & 65 / 78 \\
\cline{3-12}
\Xcline{2-12}{2.5\arrayrulewidth}
& 30 & Init. & \textbf{32} / 45 & 23 / 27 & \textbf{27} / 37 & \textbf{46} / 55 & 26 / 35 & 30 / 44 & 69 / 78 & 63 / 69 & 63 / 74 \\
\cline{3-12}
& (all psgs) & Shuf. & 30 / 45 & 20 / 32 & 24 / 38 & 37 / \textbf{58} & 24 / \textbf{41} & 31 / \textbf{53} & 69 / \textbf{80} & 58 / 74 & 65 / 77 \\
\cline{3-12}
\Xcline{2-12}{2.5\arrayrulewidth}
\Xhline{4\arrayrulewidth}
\multirow{6}{*}{\rotatebox{90}{Deep (90 psg)}}
& 1 & --  & 20 / 29 & \textbf{15} / 20 & 15 / 20 & \textbf{34} / 51 & \textbf{23} / 28 & 25 / 31 & 63 / 72 & \textbf{62} / 68 & \textbf{62} / 70 \\
\cline{3-12}
\Xcline{2-12}{2.5\arrayrulewidth}
& 30 &  Init. & 20 / 34 & 13 / 21 & \textbf{18} / 28 & 30 / 49 & 20 / 30 & \textbf{29} / 40 & \textbf{64} / 75 & 55 / 67 & 60 / 72 \\
\cline{3-12}
& (sub-batch) &  STB & 20 / \textbf{36} & 14 / \textbf{27} & 17 / \textbf{31} & 32 / \textbf{52} & 19 / 40 & 23 / 42 & \textbf{64} / \textbf{77} & 56 / \textbf{73} & 60 / \textbf{75} \\
\cline{3-12}
& & BTS  & \textbf{22} / \textbf{36} & 14 / 26 & 17 / 29 & 29 / \textbf{52} & 18 / 37 & 25 / \textbf{44} & \textbf{64} / \textbf{77} & 55 / 72 & 60 / \textbf{75} \\
\cline{3-12}
\Xcline{2-12}{2.5\arrayrulewidth}
& 90 & Init.  & 16 / 26 & 11 / 16 & 13 / 20 & 22 / 48 & 16 / 33 & 23 / 37 & 52 / 67 & 48 / 57 & 45 / 54 \\
\cline{3-12}
& (all psg) &  Shuf. & 14 / 29 & 11 / 21 & 13 / 20 & 22 / 50 & 15 / \textbf{42} & 19 / \textbf{44} & 53 / 73 & 45 / 65 & 46 / 60 \\
\cline{3-12}
\Xcline{2-12}{2.5\arrayrulewidth}
\Xhline{4\arrayrulewidth}
\end{tabular}
\caption{AUC$^m$, representing the AUC-PR at $m$ self-consistency calls/passage, for PW relevance assessment methods at $m = 1$ (no self-consistency) and $m = 15$. Increasing $m$ improves all PW methods, but the batched PW methods improve faster, becoming the best methods at $m = 15$, likely because they create more diverse self-consistency ensembles. The highest AUC-PR for each $m$ and LLM is in bold.}
\label{tab:n_scores}
\end{table*}

\subsection{RQ1: Effects of Batching and Self-Consistency on Relevance Assessment}
Figures \ref{fig:n_scores_shall} and \ref{fig:n_scores_deep} and Table \ref{tab:n_scores} show the effects of increasing the number of self-consistency LLM calls/passage ($m$) from 1 (i.e., no self-consistency) to 15 on relevance assessment performance (AUC-PR) -- with more plots in Appendix \ref{sec:appendix_n_scores_auc}. While one-by-one PW methods are competitive at $m=1$, they are always outperformed by a batched PW method at $m = 15$.

\paragraph{Batching can amplify the benefits of self-consistency:} Though increasing the number of self-consistency calls greatly improved all PW methods, batched methods improved considerably more than one-by-one methods. We conjecture that this is because batching leads to more diverse LLM sampling across self-consistency calls. For instance, for GPT-4o on Legal Search (Shallow), one-by-one PW improved from 32\% AUC-PR at $m = 1$ to 41\% at $m = 15$ ($+ 9\%$), while all-in-one PW (Shuffled) improved from 30\% to 45\% ($+ 15\%)$, respectively.

\paragraph{Batching can reduce latency by an order of magnitude or more:} Figure \ref{fig:runtimes} shows per-query runtimes across different values of $m$ and varying numbers of parallel LLMs for shallow  Legal Search. The median time for 30 sequential one-by-one PW LLM calls was 34–58s, depending on the LLM, while judging all 30 passages in a single call took only 2–6s, constituting a 6-17x speedup. Such speedups can be expected if the number of parallel LLMs is not greater than $m$ -- otherwise, smaller batch sizes allow for more parallelization, as shown in the second pane of Figure \ref{fig:runtimes}. But if there are only a few parallel LLMs available, batched methods are much faster, especially when self-consistency ensembling is used.

\paragraph{\textbf{Shuffling is helpful for high-enough $m$:}} At $m=15$, the best performance across all datasets and LLMs was always achieved by a batched method with shuffling, which can likely be attributed to more diverse LLM sampling across self-consistency calls.

\paragraph{\textbf{Sub-batching is useful for deep search:}} While all batched self-consistency methods performed well in shallow search (30 passages), for deep search (90 passages), the all-in-one methods performed poorly -- far worse than the sub-batched methods (green lines in Figure \ref{fig:n_scores_deep}). RQ2 further explores why the overly large batches of 90 passages degraded performance at the values of $m$ tested. 

\begin{figure}[!t]
    \centering
    \includegraphics[width=\linewidth]{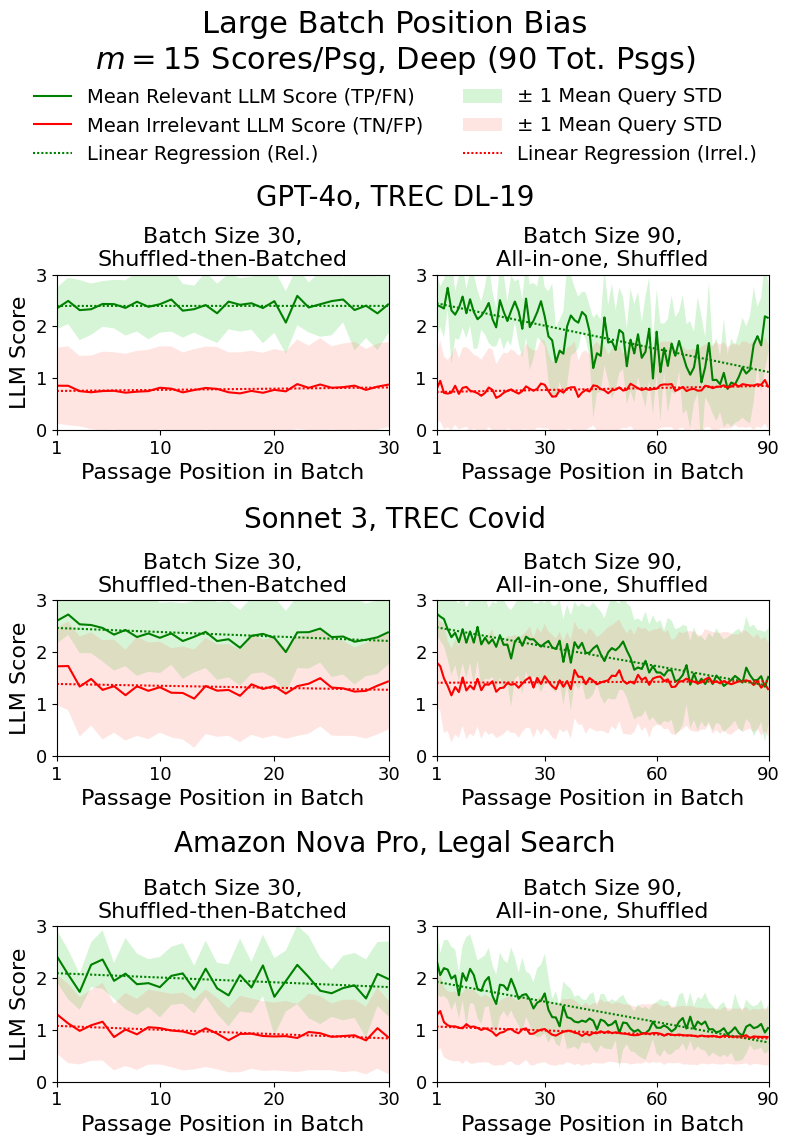}
    \caption{Mean relevant and irrelevant LLM scores with sub-batches of 30 psgs/batch (left) vs. all-in-one batches of 90 psgs/batch (right). The 30 passage batches are relatively consistent in discriminating relevance throughout the batch, while the 90 passage batches lose most of their discriminative power towards the tail of the batch.}
    \label{fig:pos_bias}
\end{figure}

\subsection{RQ2: Effects of Position Biases}
Before considering the effect of the initial passage order (RQ3), we first ask whether batched LLM scoring exhibits consistent positional biases across \textit{random} permutations of a given passage list. Figure \ref{fig:pos_bias} shows LLM scores versus passage positions in a batch, with each passage seen by an LLM in $m = 15$ random permutations.

\paragraph{Large batches have harmful position biases that can be mitigated by sub-batching:} The harmful biases in the large 90 passage batches in Figure \ref{fig:pos_bias} are obvious: the capacity to discriminate between relevant vs. non-relevant passages is almost gone towards the tail of the batch, with GPT-4o showing a clear lost-in-the-middle effect \cite{liu2024lost}. By comparison, sub-batching with 30 passages per batch is far more consistent in being able to discriminate relevance throughout the batch, explaining its far superior performance on deep search. 

\subsection{RQ3: Effects of Initial Passage Order}

\begin{figure}[!b]
    \centering
    \includegraphics[width=\linewidth]{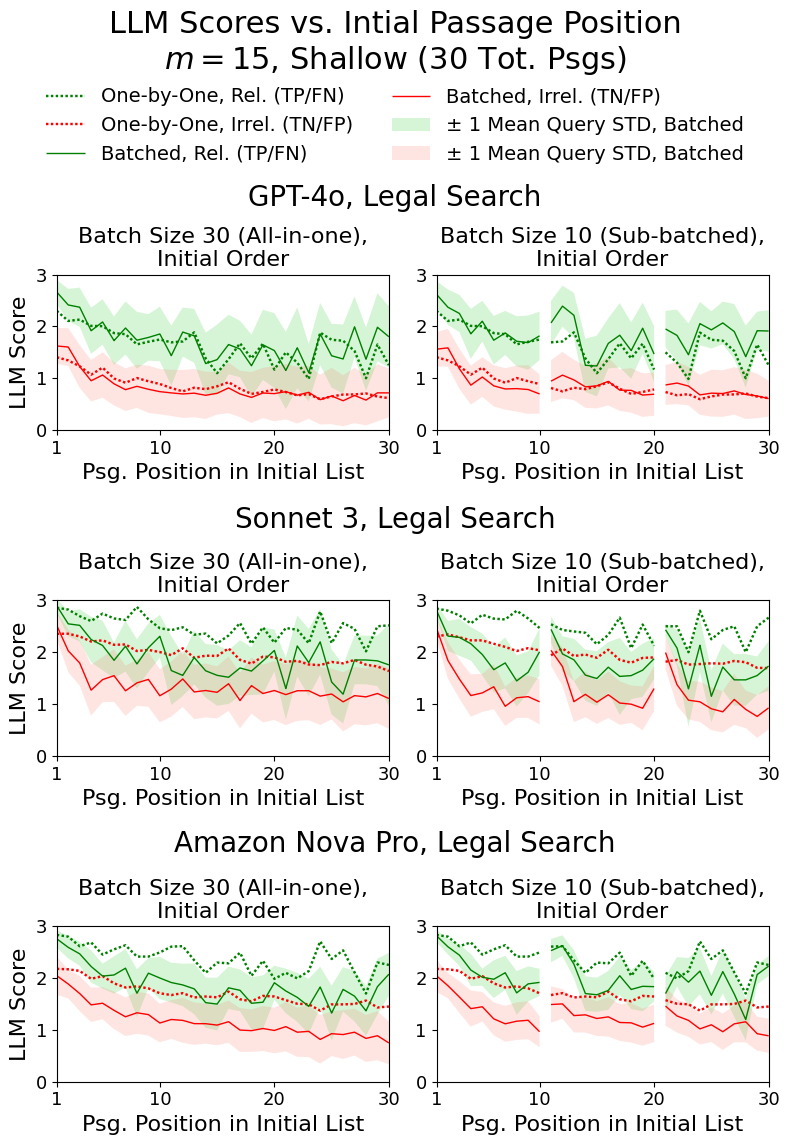}
    \caption{Mean LLM scores for one-by-one PW versus batched PW methods with initial order. For all-in-one PW methods (left), GPT-4o tracks much more closely to one-by-one PW than Sonnet and Nova Pro. Sub-batching with initial order (right) creates artificial cycles and discontinuities at batch junctions.}
    \label{fig:discont}
\end{figure}

Next we investigate the potential positional biases caused by the initial list order $L^q$. Figure \ref{fig:discont} compares several batched methods that use the order of $L^q$ versus one-by-one scoring, which does not depend on the order of $L^q$.

\paragraph{GPT-4o has the least batching bias:} For GPT 4o all-in-one PW in Figure \ref{fig:discont}, the batched scores track quite closely to one-by-one PW scores, though the front and tail of the batch have slightly higher scores. In contrast, batched Sonnet and Nova Pro scores are typically far lower than their respective one-by-one scores everywhere except at the very front of the batch, helping explain the weakness of these LLMs.

\paragraph{Sub-batching with initial order can induce cycles and discontinuities} The RHS of Figure \ref{fig:discont} shows that sub-batching with initial order can cause score peaks at the start of every sub-batch -- creating discontinuities at batch junctions and inducing cyclical score fluctuations. This explains why sub-batching with the initial order performs worse than the shuffling variants.

\begin{table*}[!ht]
\scriptsize
\centering
\begin{tabular}{|c|c|c|c||c|c|c||c|c|c||c|c|c||}
\cline{1-13}
 & \textbf{Psgs/} & & \textbf{Psg} & \multicolumn{3}{c||}{\textbf{Legal Search}} & \multicolumn{3}{c||}{\textbf{DL-19}} & \multicolumn{3}{c||}{\textbf{Covid}} \\
\cline{5-13}
 \textbf{Method} & \textbf{Batch} & $m$ & \textbf{Order} & \textbf{GPT-4o} & \textbf{Sonnet} & \textbf{Nova} & \textbf{GPT-4o} & \textbf{Sonnet} & \textbf{Nova} & \textbf{GPT-4o} & \textbf{Sonnet} & \textbf{Nova} \\
\Xhline{4\arrayrulewidth}
 Initial & -- & -- & -- & 37.2 & 37.2 & 37.2 & 50.6 & 50.6 & 50.6 & 59.5 & 59.5 & 59.5 \\
 \Xcline{1-13}{3\arrayrulewidth} 
 \multirow{4}{*}{LW} & \multirow{4}{*}{30} & \multirow{2}{*}{1} & Init. & 46.2 & 41.6 & 44.5 & 65.9 & 65.2 & 61.9 & 73.4 & 66.4 & 67.9 \\
 \cline{4-13}
   & & & Shuf. & 13.9 & 12.6 & 13.8 & 34.6 & 36.0 & 34.6 & 45.7 & 47.5 & 48.3 \\
  \Xcline{3-13}{3\arrayrulewidth} 
  &  & \multirow{2}{*}{15} & Init. & 48.9 & 45.1 & 46.7 & 67.4 & 66.7 & 63.6 & 74.6 & 70.5 & 70.3 \\
 \cline{4-13}
  &  &  & Shuf. & 13.9 & 15.3 & 13.3 & 33.0 & 30.4 & 30.8 & 45.0 & 47.4 & 45.2 \\
 \Xcline{1-13}{3\arrayrulewidth} 
 \multirow{12}{*}{PW} & \multirow{2}{*}{1} & 1 & N/A & 45.6 & 42.7 & 41.7 & 63.3 & 63.4 & 64.1 & 75.2 & 73.9 & 75.7 \\
 \cline{3-13}
  &  & 15 & N/A & 46.5 & 44.5 & 43.0 & 67.8 & 65.4 & 64.7 & 76.6 & 76.0 & 78.8 \\
 \Xcline{2-13}{3\arrayrulewidth} 
  & \multirow{6}{*}{10} & \multirow{3}{*}{1} & Init. & 45.5 & 40.2 & 42.4 & 65.6 & 63.5 & 65.9 & 75.6 & 72.7 & 75.8 \\
  \cline{4-13}
  & & & STB & 45.5 & 38.5 & 42.8 & 67.5 & 63.0 & 64.8 & 74.7 & 74.6 & 75.0\\
  \cline{4-13}
 & & & BTS & 45.3 & 39.2 & 42.1 & 66.6 & 63.0 & 67.0 & 75.4 & 73.3 & 75.5\\
  \Xcline{3-13}{3\arrayrulewidth} 
 & & \multirow{3}{*}{15} & Init. & 48.5 & 38.9 & 43.6 & 67.8 & 64.9 & 67.4 & 77.9 & 74.2 & 76.6 \\
 \cline{4-13}
  &  &  & STB & \textbf{50.0} & \textbf{45.4} & 48.2 & 68.6 & \textbf{67.8} & 68.0 & \textbf{80.7} & \textbf{79.6} & \textbf{80.4} \\
 \cline{4-13}
  &  &  & BTS & 48.1 & 43.1 & 45.4 & 68.9 & 67.0 & \textbf{68.8} & 78.7 & 79.5 & 79.6 \\
 \Xcline{2-13}{3\arrayrulewidth} 
  & \multirow{4}{*}{30} & \multirow{2}{*}{1} & Init. & 46.1 & 42.8 & 43.6 & 66.5 & 63.4 & 65.6 & 76.1 & 73.2 & 73.6 \\
  \cline{4-13}
  & & & Shuf. & 44.5 & 34.8 & 43.9 & 66.7 & 64.1 & 64.3 & 75.6 & 69.5 & 73.3 \\
  \Xcline{3-13}{3\arrayrulewidth} 
& & \multirow{2}{*}{15} & Init. & 49.4 & 41.8 & 46.9 & \textbf{69.3} & 65.6 & 68.0 & 77.9 & 74.3 & 76.7 \\
  \cline{4-13}
  &  &  & Shuf. & \textbf{50.0} & 41.3 & \textbf{48.3} & 68.6 & 67.7 & 67.6 & 80.3 & 77.2 & 79.1 \\
\Xcline{1-13}{3\arrayrulewidth}
\Xhline{4\arrayrulewidth}
\end{tabular}
\caption{NDCG@10 (\%) for shallow reranking (30 total passages) for all LW and PW LLM methods with $m = 1$ (i.e., no self-consistency) $m = 15$ (i.e., 15 self-consistency calls/psg). LW and one-by-one PW methods are competitive at $m = 1$ but do not benefit as strongly from self-consistency as batched PW methods, causing the later to achieve the best NDCG@10 at $m = 15$. }
\label{tab:ndcg_shall}
\end{table*}

\begin{table*}[!ht]
\scriptsize
\centering
\begin{tabular}{|c|c|c|c||c|c|c||c|c|c||c|c|c||}
\cline{1-13}
 & \textbf{Psgs/} & & \textbf{Psg} & \multicolumn{3}{c||}{\textbf{Legal Search}} & \multicolumn{3}{c||}{\textbf{DL-19}} & \multicolumn{3}{c||}{\textbf{Covid}} \\
\cline{5-13}
 \textbf{Method} & \textbf{Batch} & $m$ & \textbf{Order} & \textbf{GPT-4o} & \textbf{Sonnet} & \textbf{Nova} & \textbf{GPT-4o} & \textbf{Sonnet} & \textbf{Nova} & \textbf{GPT-4o} & \textbf{Sonnet} & \textbf{Nova} \\
\Xhline{4\arrayrulewidth}
 Initial & -- & -- & -- & 37.2 & 37.2 & 37.2 & 50.6 & 50.6 & 50.6 & 59.5 & 59.5 & 59.5 \\
 \Xcline{1-13}{3\arrayrulewidth} 
 \multirow{4}{*}{LW} & \multirow{4}{*}{90} & \multirow{2}{*}{1} & Init. & 46.0 & 41.1 & 36.4 & 70.3 & 64.6 & 56.1 & 76.2 & 67.1 & 64.1 \\
 \cline{4-13}
  & & & Shuf. & 9.4 & 8.6 & 9.4 & 23.5 & 25.5 & 23.8 & 42.0 & 39.4 & 39.5 \\
   \Xcline{3-13}{3\arrayrulewidth} 
  &  & \multirow{2}{*}{15} & Init. & 48.4 & 42.5 & 42.4 & 72.5 & 66.7 & 67.2 & 65.2 & 60.1 & 63.3 \\
 \cline{4-13}
  &  &  & Shuf. & 13.2 & 12.7 & 12.8 & 26.3 & 25.4 & 25.7 & 44.5 & 45.2 & 44.3 \\
 \Xcline{1-13}{3\arrayrulewidth} 
 \multirow{12}{*}{PW} & \multirow{2}{*}{1} & 1 & N/A & 44.9 & 41.8 & 41.7 & 69.8 & 66.4 & 67.7 & 78.9 & 77.8 & 80.0 \\
 \cline{3-13}
  &  & 15 & N/A & 46.8 & \textbf{43.5} & 42.9 & 73.6 & 68.6 & 69.3 & 80.1 & 79.8 & 83.3 \\
 \Xcline{2-13}{3\arrayrulewidth} 
  & \multirow{6}{*}{30} & \multirow{3}{*}{1} & Init. & 43.6 & 38.2 & 42.8 & 72.3 & 62.0 & 68.7 & 82.9 & 72.2 & 75.6 \\
  \cline{4-13}
  & & & STB & 43.8 & 34.9 & 44.7 & 70.7 & 63.2 & 68.7 & 79.8 & 72.6 & 78.2 \\
  \cline{4-13}
 & & & BTS &  40.4 & 33.7 & 42.7 & 69.5 & 64.0 & 67.6 & 81.6 & 71.6 & 75.2 \\
  \Xcline{3-13}{3\arrayrulewidth} 
 & & \multirow{3}{*}{15} & Init. & 47.1 & 36.3 & 44.1 & 71.0 & 66.0 & 69.5 & 83.8 & 77.4 & 79.9 \\
 \cline{4-13}
  &  &  & STB & \textbf{51.3} & 41.3 & \textbf{49.8} & \textbf{76.3} & \textbf{71.9} & \textbf{72.1} & 86.1 & 82.1 & \textbf{84.3} \\
 \cline{4-13}
  &  &  & BTS & 50.6 & 39.2 & 46.7 & 73.9 & 70.0 & 70.2 & \textbf{86.3} & \textbf{82.5} & 83.5 \\
 \Xcline{2-13}{3\arrayrulewidth} 
  & \multirow{4}{*}{90} & \multirow{2}{*}{1} & Init. & 43.5 & 37.7 & 41.9 & 66.9 & 62.2 & 64.4 & 73.8 & 71.8 & 68.6 \\
  \cline{4-13}
  & & & Shuf. & 29.0 & 26.4 & 32.7 & 55.4 & 52.5 & 51.3 & 63.3 & 61.3 & 59.0 \\
  \Xcline{3-13}{3\arrayrulewidth} 
& & \multirow{2}{*}{15} & Init. & 48.5 & 39.7 & 45.0 & 72.1 & 66.7 & 66.3 & 79.3 & 77.0 & 72.5 \\
  \cline{4-13}
  &  &  & Shuf. & 45.1 & 28.4 & 40.4 & 73.0 & 61.2 & 65.9 & 82.6 & 71.8 & 74.4 \\
\Xcline{1-13}{3\arrayrulewidth}
\Xhline{4\arrayrulewidth}
\end{tabular}
\caption{NDCG@10 (\%) for deep reranking (90 total passages) for all LW and PW methods at $m \in \{1, 15\}$. Sub-batched methods (30 psg/batch) perform best at $m = 15$ with the STB variant typically achieving the highest NDCG@10, likely due having the most diverse batching strategy and avoiding large-batch position biases.}
\label{tab:ndcg_deep}
\end{table*}


\subsection{RQ4: Ranking Performance of Batched Self Consistency Methods}
Ranking performance in terms of NDCG@10 for all LLM methods is shown in Tables \ref{tab:ndcg_shall} and \ref{tab:ndcg_deep} for shallow and deep search, respectively, with  detailed results on the effects of $m$ in Appendix \ref{sec:appendix_n_scores_ndcg_10}. 

\paragraph{Batching amplifies self-consistency benefits for ranking:} Without self consistency ($m = 1$), one-by-one PW and LW (initial order) methods are competitive rankers, but adding self-consistency helps batched PW methods more than it helps these baselines -- making the batched PW methods with $m=15$ the strongest rankers overall.  
For instance, as seen in Table \ref{tab:ndcg_deep} for Legal Search (deep), GPT-4o one-by-one PW ranking improves from 44.9\% NDCG@10 with $m=1$ to 46.8\% with $m=15$, while sub-batched (STB) PW ranking improves from 43.8\% to 51.3\%, respectively.

\paragraph{STB ($m=15$) performs best:} Sub-batched STB methods with $m=15$ performed best overall, likely due to creating the broadest range of contexts for LLM sampling by creating the most diverse passage permutations and subsets. All-in-one batched PW methods (which needed 3 times fewer LLM calls than STB) with $m = 15$ were also effective for shallow search (30 passages), but under-performed for deep search (90 passages), likely due to the harmful large-batch biases seen in Figure \ref{fig:pos_bias}.

\paragraph{LW ranking with the initial list order is competitive with one-by-one PW ranking:} When the initial list order $L^q$ is kept, LW ranking is competitive with one-by-one PW ranking, but when $L^q$ is shuffled, LW methods perform very poorly.


\section{Conclusion}
We show that batched PW methods for passage relevance assessment and ranking can improve not only efficiency, but also judgment quality by enabling content from multiple passages to be seen jointly. Further, when self-consistency ensembles are used to collect and aggregate multiple scores per passage, batched methods can create more diversity between ensemble components than one-by-one methods. While score variation in one-by-one methods comes only from  LLM stochasticity, batching can naturally diversify the contexts in which a passage is scored through different co-candidate subsets and permutations -- with autoregressive multi-score sequence generation diversifying scores even further. As our experiments show, this leads to batching amplifying the test-time scaling benefits of self-consistency, giving batched PW methods the best performance while achieving order-of-magnitude level speedups over one-by-one PW methods.

\section*{Limitations}
The main limitation of our work are the computational resources required, since LLM relevance assessment and ranking is expensive computationally. We thus only tested a range of $m \in \{1,\cdots,15\}$ for the number of self-consistency calls, even though higher levels of $m$ would have added information to our results. Also due to computational limitations, we only tested three LLMs (GPT-4o, Claude Sonnet 3, and Amazon Nova Pro) across three datasets (TREC DL-19, TREC Covid, and Legal Search), though it would be interesting to test an even wider range of models and datasets. Similarly, we were limited to testing four batch sizes $\in \{1,10,30,90\}$ across five batching strategies (c.f. Figure \ref{fig:main}) and two levels of search: shallow (30 initial passages) and deep (90 initial passages).

As another limitation, we note that LLMs likely will have seen open-source datasets such as TREC DL-19 and TREC Covid during pretraining, which is why using the third, closed-source Legal Search dataset is very important in our experiments. Fortunately, we are able to observe that our results generalize across both the open-source and closed-source data.

Finally, we must point out several risks of using LLMs for ranking and relevance assessment at scale. Firstly, LLMs can amplify societal biases that they will have learned during their pretraining process, creating a risk for harm. Secondly, LLMs carry a risk of ``jail-breaking", or malicious prompt injection, creating safety risks. Finally, LLMs may provide incorrect judgments on passage relevance, which could have severely negative effects for high-stakes applications.

\bibliography{custom}

\appendix

\section{Relevance Assessment Quality vs Scores per Passage}
\label{sec:appendix_n_scores_auc}

Figures \ref{fig:shallow_auc_grid} and \ref{fig:deep_auc_grid} below show the effects of $m$ on AUC-PR of one-by-one PW and batched PW methods for all datasets and LLMs for shallow and deep search, respectively.

\begin{figure*}[!htbpp]
    \centering
    \includegraphics[width = \linewidth]{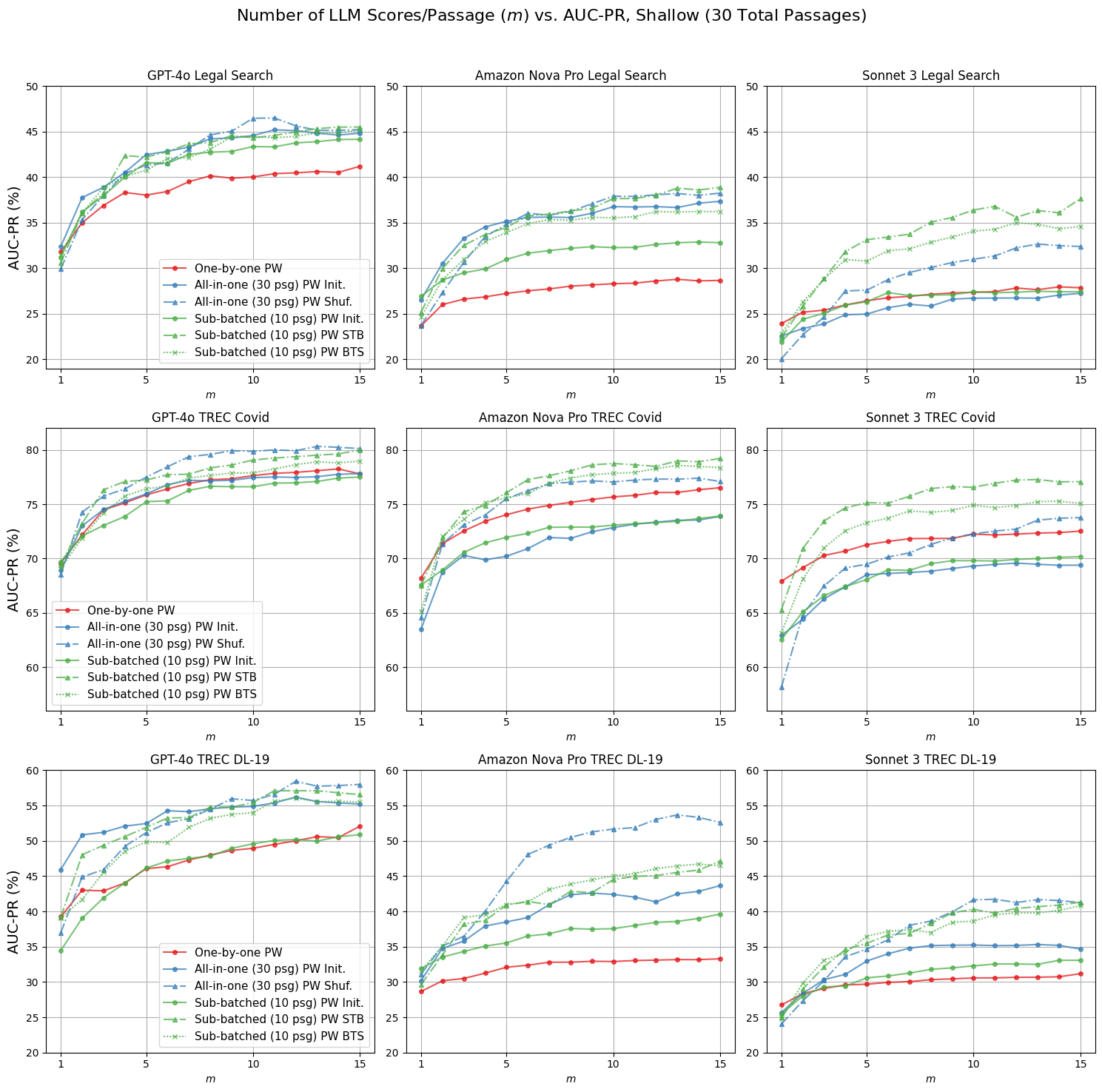}
    \caption{Number of LLM Scores/Passage ($m$) vs. AUC-PR, Shallow (30 Total Passages)}
    \label{fig:shallow_auc_grid}
\end{figure*}

\begin{figure*}[!htbp]
    \centering
    \includegraphics[width = \linewidth]{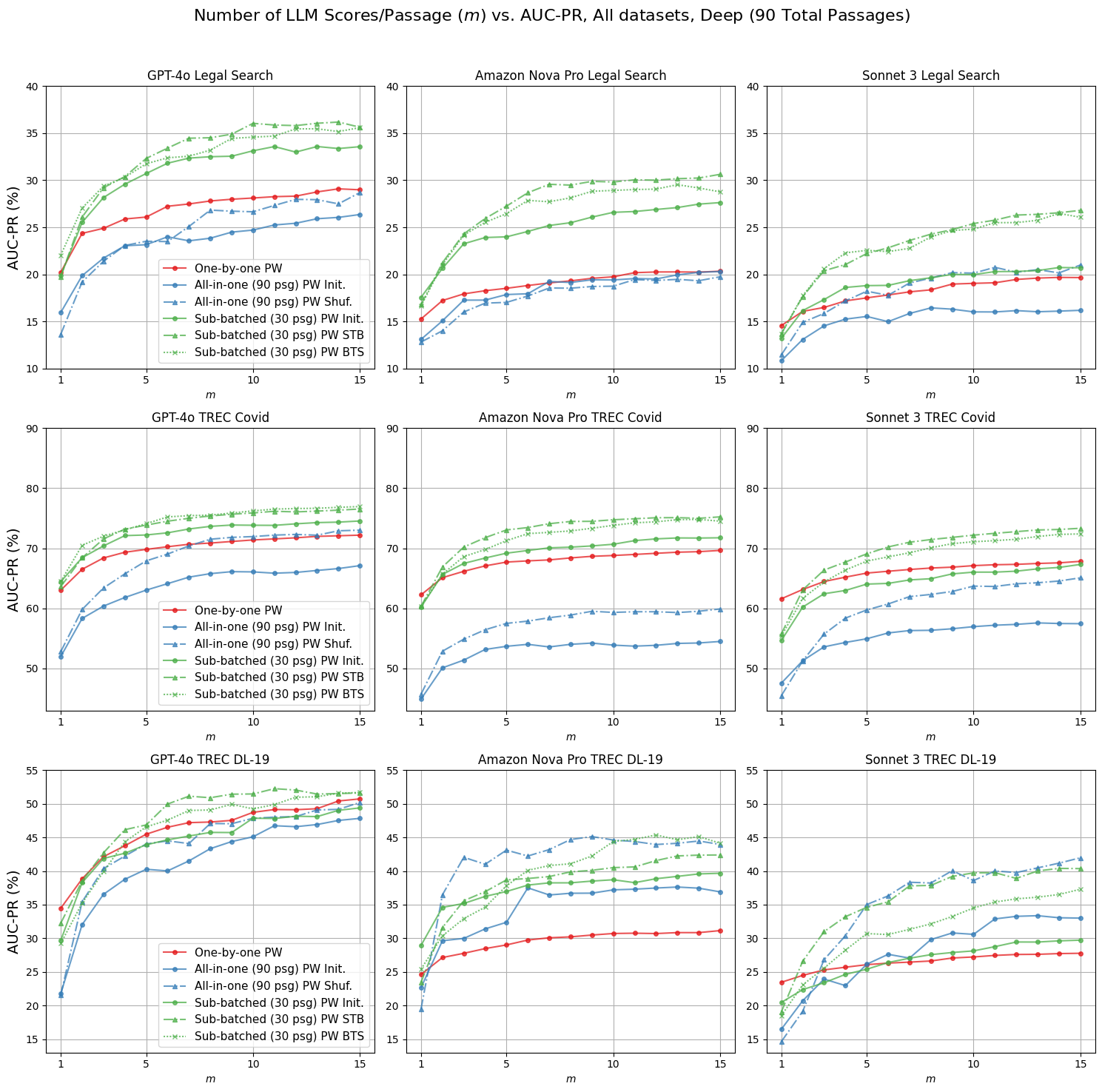}
    \caption{Number of LLM Scores/Passage ($m$) vs. AUC-PR, Deep (90 Total Passages)}
    \label{fig:deep_auc_grid}
\end{figure*}

\section{NDCG@10 vs Scores per Passage}
\label{sec:appendix_n_scores_ndcg_10}

Figures \ref{fig:shallow_ndcg_grid} and \ref{fig:deep_ndcg_grid} below show the effects of $m$ on NDCG@10 of one-by-one PW and batched PW methods for all datasets and LLMs for shallow and deep search, respectively.

\section{Experiment Runtimes}
\begin{figure*}[!htbpp]
    \centering
    \includegraphics[width=\linewidth]{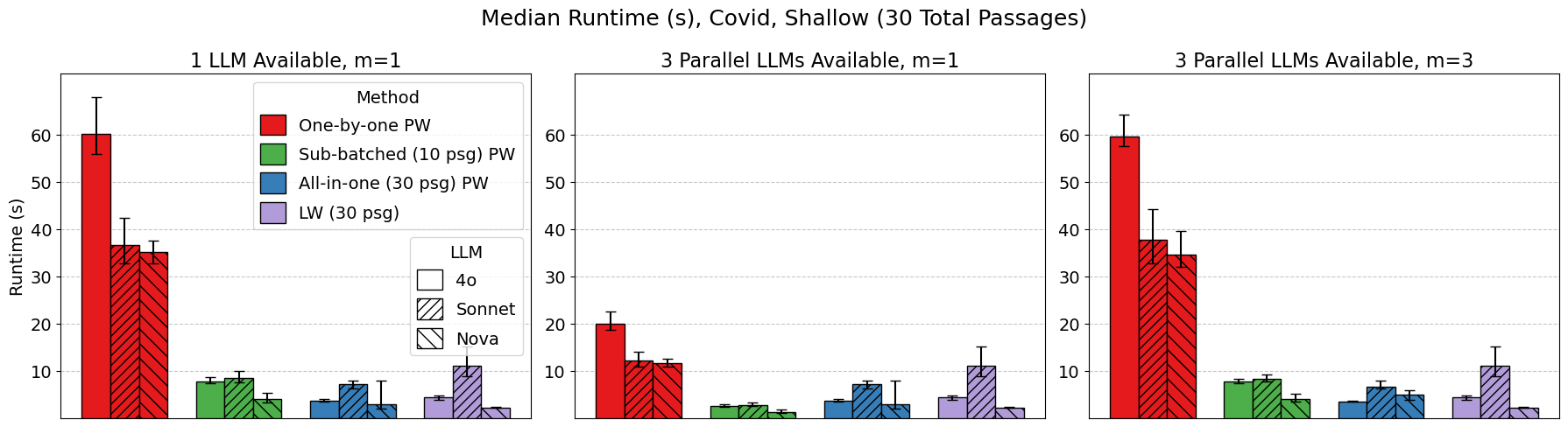}
    \caption{Median per-query runtimes for Covid, Shallow (30 total passages), with error bars showing the IQR.}
\label{fig:runtimes_covid_shal}
\end{figure*}

\begin{figure*}[!htbpp]
    \centering
    \includegraphics[width=\linewidth]{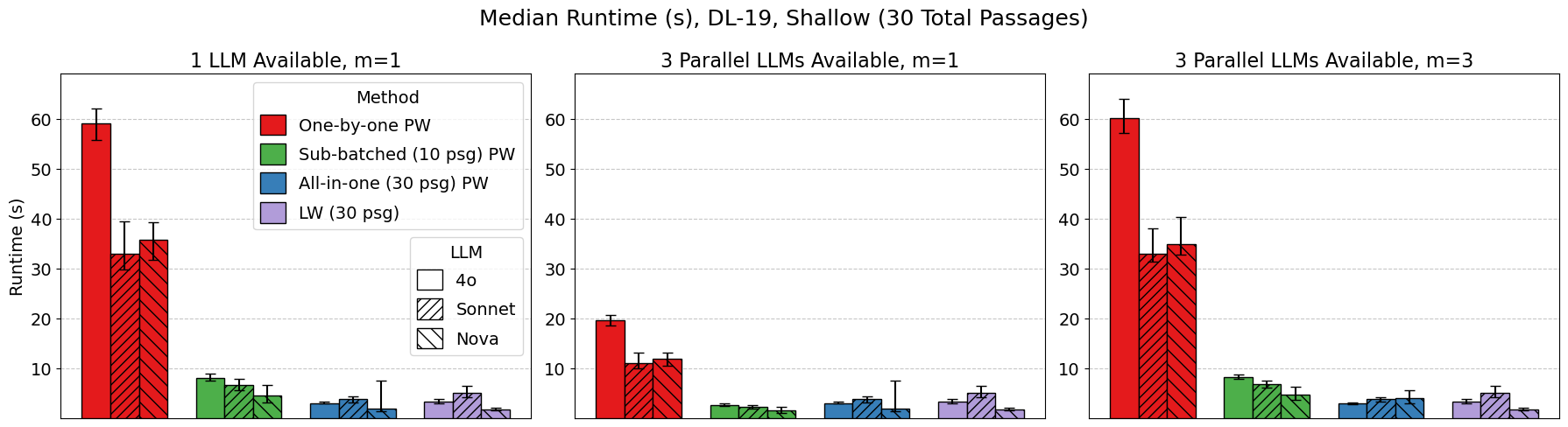}
    \caption{Median per-query runtimes for DL-19 Search, Shallow (30 total passages), with error bars showing the IQR.}
    \label{fig:runtimes_dl19_shal}
\end{figure*}

\begin{figure*}[!htbpp]
    \centering
    \includegraphics[width=\linewidth]{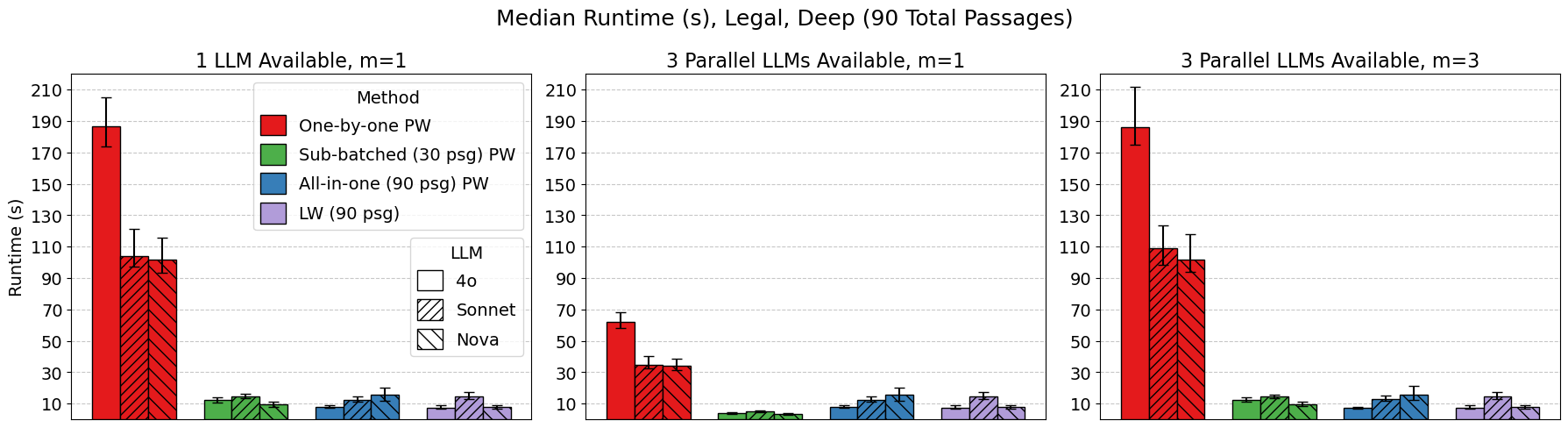}
    \caption{Median per-query runtimes for Legal Search, Deep (90 total passages), with error bars showing the IQR.}
\label{fig:runtimes_legal_deep}
\end{figure*}

\begin{figure*}
    \centering
    \includegraphics[width=\linewidth]{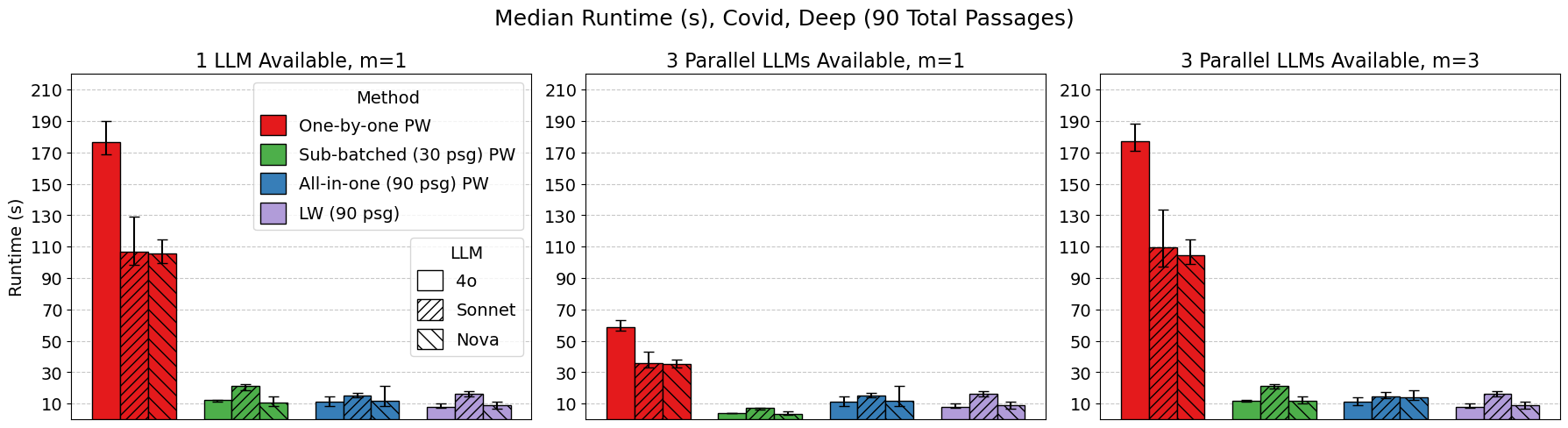}
    \caption{Median per-query runtimes for Covid, Deep (90 total passages), with error bars showing the IQR.}
    \label{fig:runtimes_covid_deep}
\end{figure*}

\begin{figure*}
    \centering
    \includegraphics[width=\linewidth]{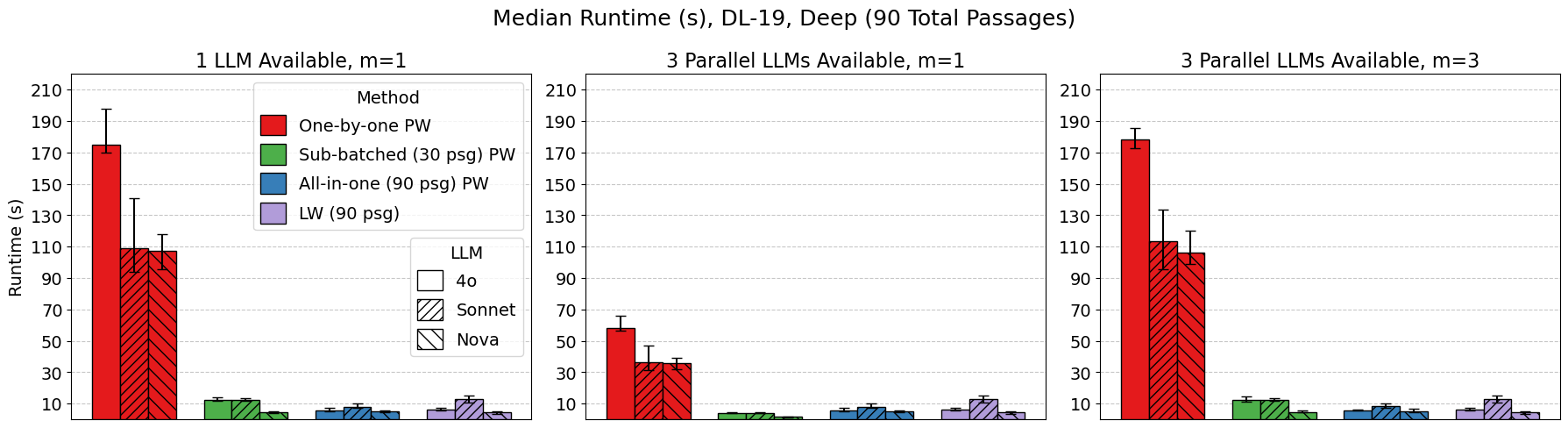}
    \caption{Median per-query runtimes for DL-19, Deep (90 total passages), with error bars showing the IQR.}
    \label{fig:runtimes_dl19_deep}
\end{figure*}

Figures \ref{fig:runtimes_covid_shal}-\ref{fig:runtimes_dl19_deep} show the per-query runtimes for all LLM methods for several values of $m$ and numbers of parallel LLMs available. 

\section{Prompt Templates}
\begin{figure}[!htb]
    \centering
    \includegraphics[width=\linewidth]{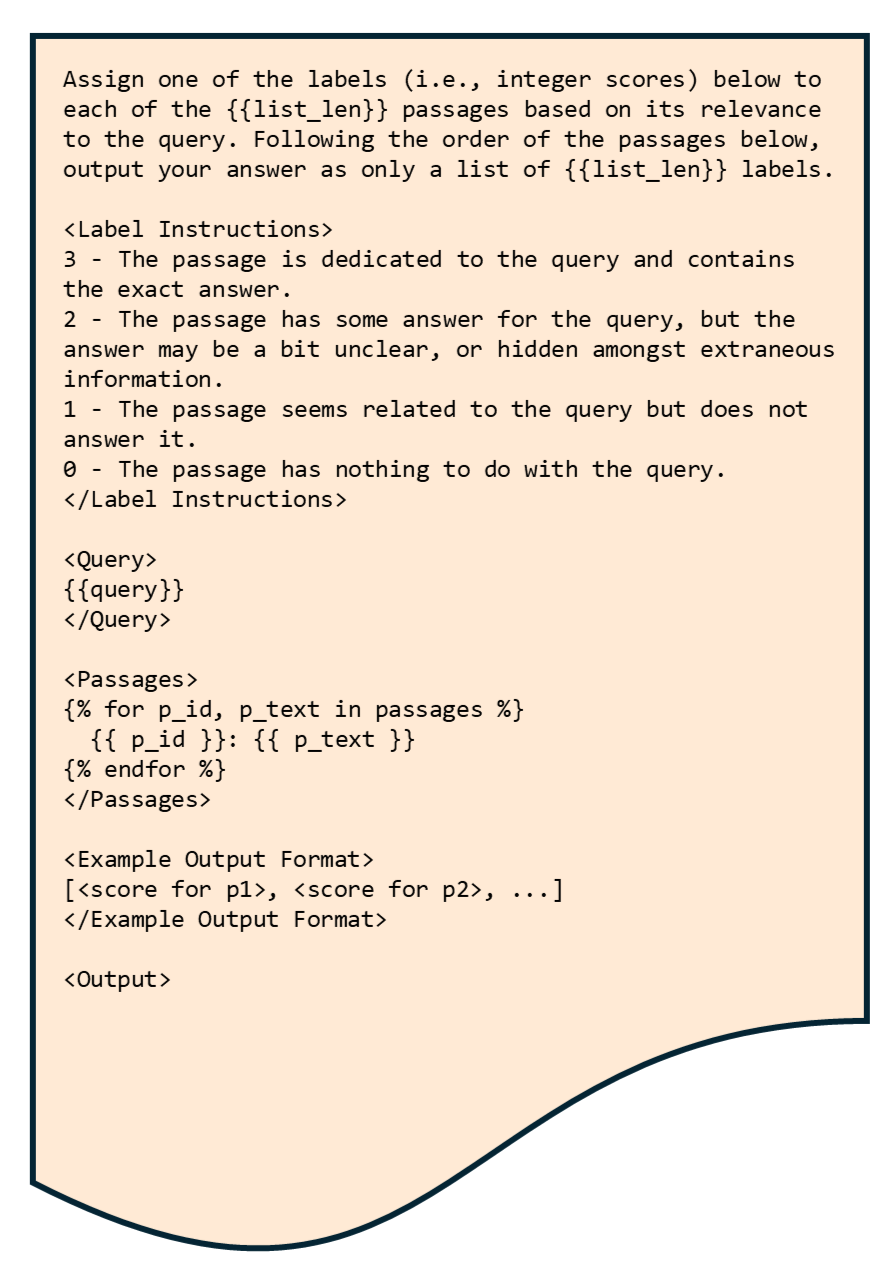}
    \caption{The pointwise relevance assessment prompt, based on the relevance label instructions from the UMBRELLA open source reproduction of the Bing relevance assessment prompt \cite{upadhyay2024umbrela}. }
    \label{fig:umb}
\end{figure}

\begin{figure}[!htb]
    \centering
    \includegraphics[width=\linewidth]{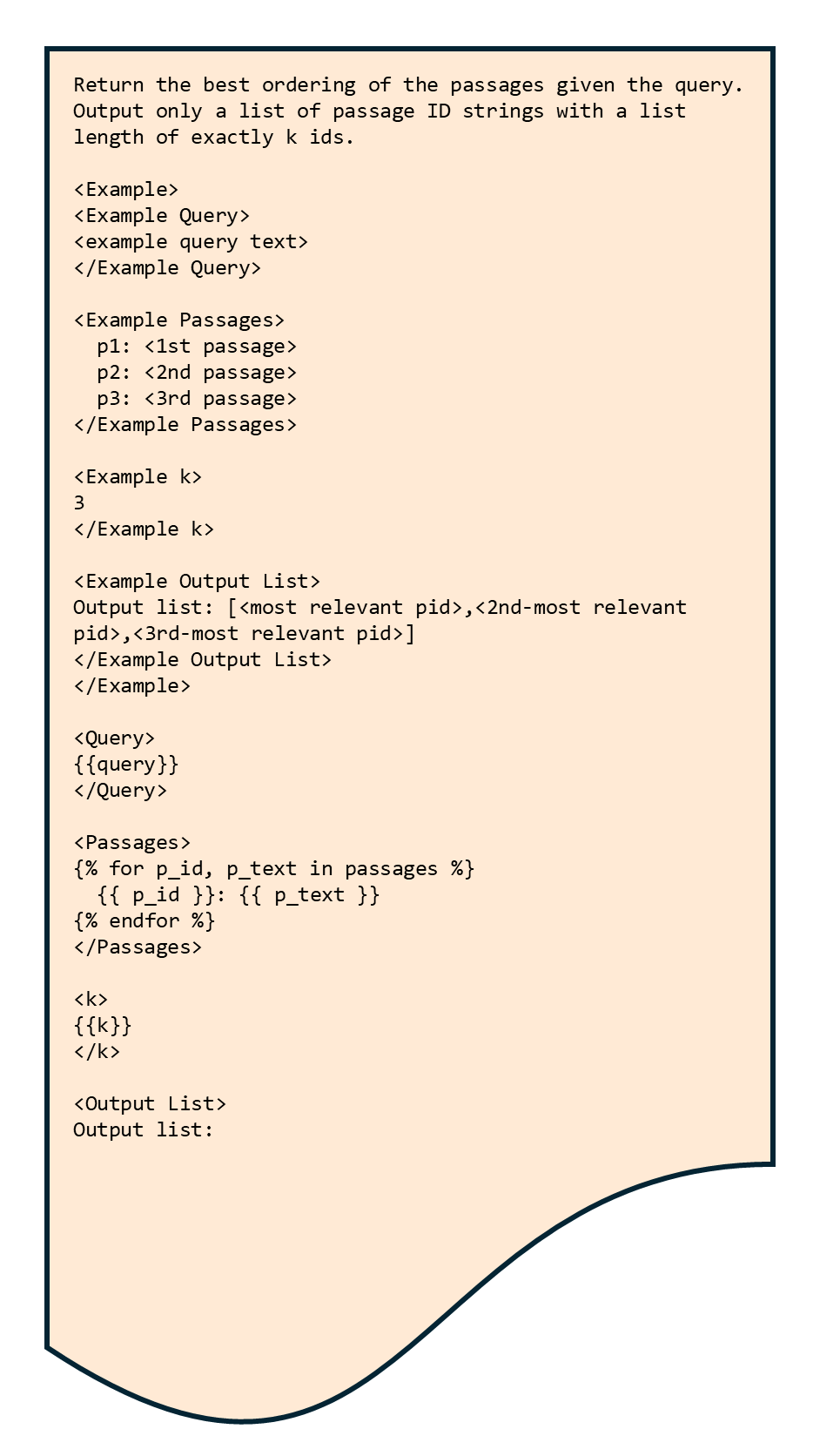}
    \caption{The listwise ranking prompt.}
    \label{fig:lw_prompt}
\end{figure}

Figures \ref{fig:umb} and \ref{fig:lw_prompt} show the full prompts used for our PW and LW implementations.

\begin{figure*}[!htbp]
    \centering
    \includegraphics[width = \linewidth]{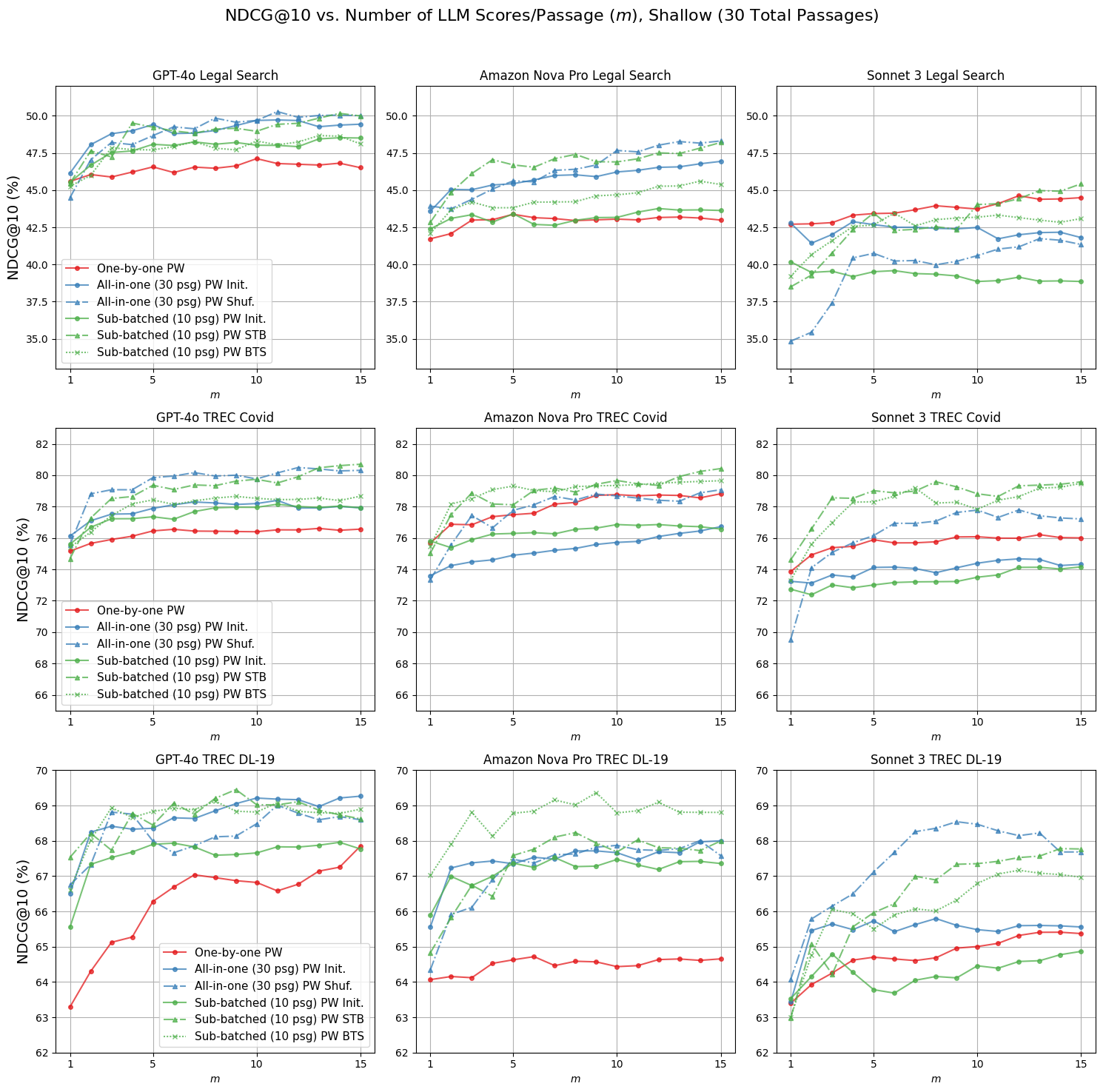}
    \caption{Number of LLM Scores/Passage ($m$) vs. NDCG@10, Shallow (30 Total Passages)}
    \label{fig:shallow_ndcg_grid}
\end{figure*}

\begin{figure*}[!htbpp]
    \centering
    \includegraphics[width = \linewidth]{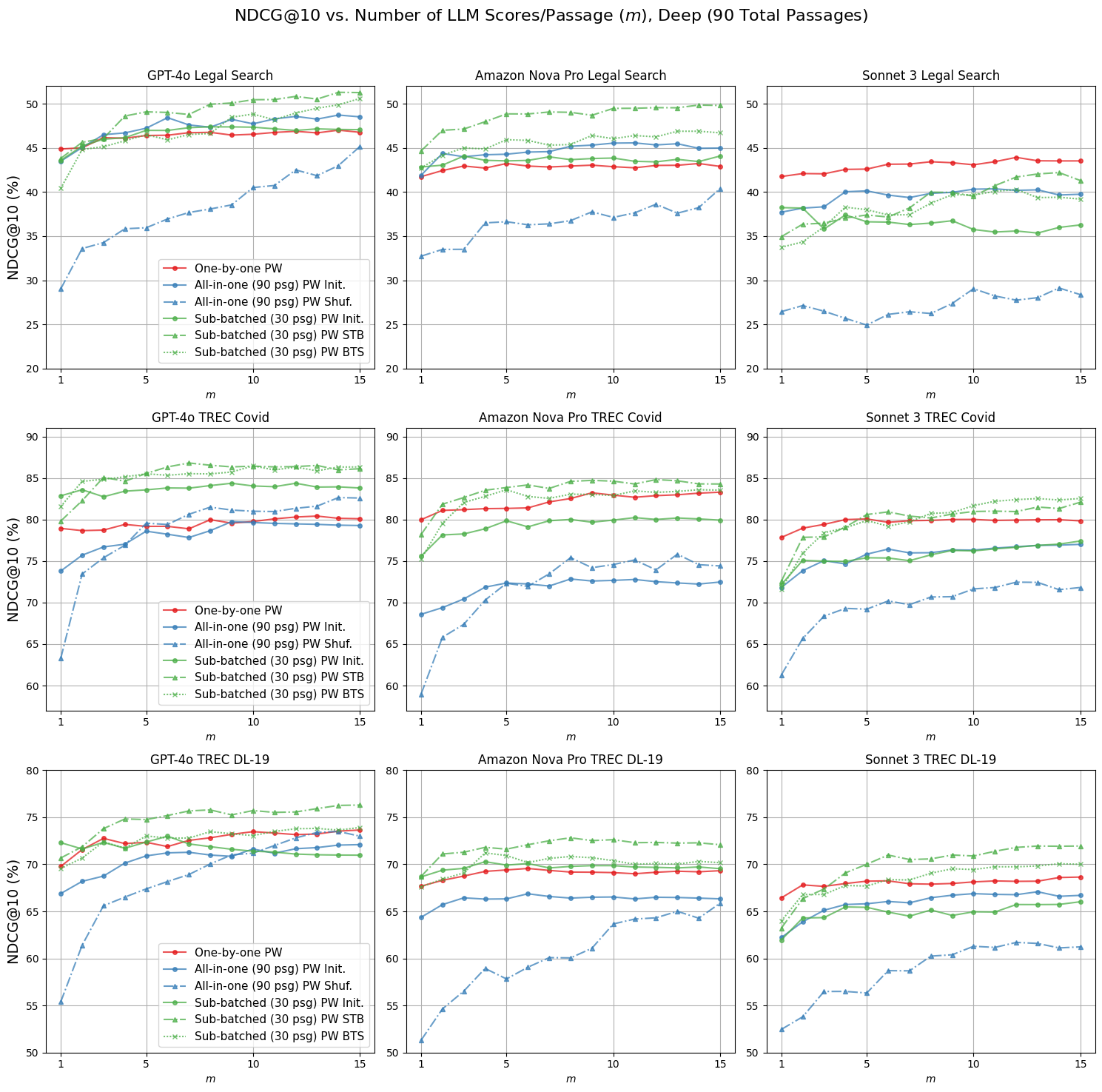}
    \caption{Number of LLM Scores/Passage ($m$) vs.NDCG@10, Deep (90 Total Passages)}
    \label{fig:deep_ndcg_grid}
\end{figure*}


\label{sec:appendix_prompts}

\end{document}